\documentclass[aps, prd, twocolumn,nofootinbib,groupedaddress,floatfix]{revtex4-2}
\usepackage{amsmath, amssymb, amsthm, float,mathtools}
\usepackage[urlcolor=blue,colorlinks,breaklinks=true]{hyperref}

\PassOptionsToPackage{obeyspaces,spaces,hyphens}{url}
\usepackage{hyperref}

\begin{document}

\title{Biased Domain Walls and the Origin of Early Massive Structures}

\author{C.~Winckler}
\email[Electronic address:\ ]{clara.winckler@astro.up.pt}
\affiliation{Departamento de Física e Astronomia, Faculdade de Ciências,
Universidade do Porto, Rua do Campo Alegre 687, PT4169-007 Porto, Portugal and\\
Instituto de Astrofísica e Ciências do Espaço, Universidade do Porto,
CAUP, Rua das Estrelas, PT4150-762 Porto, Portugal}

\author{P.~P.~Avelino}
\email[Electronic address:\ ]{pedro.avelino@astro.up.pt}
\affiliation{Departamento de Física e Astronomia, Faculdade de Ciências,
Universidade do Porto, Rua do Campo Alegre 687, PT4169-007 Porto, Portugal\\
Instituto de Astrofísica e Ciências do Espaço, Universidade do Porto,
CAUP, Rua das Estrelas, PT4150-762 Porto, Portugal}

\author{L.~Sousa}
\email[Electronic address:\ ]{lara.sousa@astro.up.pt}
\affiliation{Instituto de Astrofísica e Ciências do Espaço, Universidade do Porto,
CAUP, Rua das Estrelas, PT4150-762 Porto, Portugal and\\
Departamento de Física e Astronomia, Faculdade de Ciências,
Universidade do Porto, Rua do Campo Alegre 687, PT4169-007 Porto, Portugal}

\date{\today}

\begin{abstract}
Discrete symmetry-breaking phase transitions in the early universe may have caused the formation of networks of sheet-like topological defects, usually referred to as domain walls, which separate regions that have settled into different vacuum states. Field theory simulations predict the successive collapse of increasingly larger domains, which could potentially leave observable imprints in present-day large-scale structures. We use a non-parametric analytical model to provide an estimate of the final decay energy of these walls and their associated collapse rate, as a function of redshift. The energy released by collapsing walls can act as a seed for density perturbations in the background matter field, influencing structure formation. We estimate the dependence of the current mass of the resulting non-linear objects on the collapse redshift and wall tension, showing that domain walls can contribute to the formation of objects as massive as present-day galaxy clusters.~Still, we confirm that the contribution of standard domain walls to structure formation is subdominant. In contrast, biased domain walls --- originating in models with an approximate (or biased) discrete symmetry breaking --- generally face much less stringent constraints on their tension, which allows for significantly higher collapse energies. Based on our analysis, we are able to show that the collapse of such biased wall networks can provide a significant contribution to structure formation, and, in particular, a mass excess at $z \gtrsim 7$ as suggested by JWST data.
\end{abstract}

\maketitle

\section{Introduction}
Many particle physics scenarios predict a series of symmetry-breaking phase transitions in the early universe in which non-trivial field configurations known as topological defects may form~\cite{Kibble_1976,VilenkinBook}. When a discrete symmetry is broken, this gives rise to sheet-like defects known as domain walls which separate regions of the universe that have different vacuum expectation values. 

Domain walls tend to dominate the cosmic energy budget, since their energy density dilutes more slowly than the background energy density in a (decelerating) expanding background. As a result, cosmologically viable domain wall scenarios are tightly constrained. If domain walls were to be stable and to survive until the present time, they would have to be extremely light --- with a tension $\sigma_{\rm w}<\sigma_{\rm Zel}$, where $\sigma_{\rm Zel}=(1\ {\rm MeV})^3$ is known as the Zel'dovich bound~\cite{Zeldovich:1974uw} --- in order not to leave strong signatures on the cosmic microwave background (see also~\cite{Sousa:2015cqa} for a more detailed study that resulted in a slightly more stringent bound). This restriction may however be evaded if domain walls are unstable and the network decays early enough in cosmic history~\cite{Larsson:1996sp}. These so-called biased domain walls are, in fact, predicted in many relevant particle physics scenarios~\cite{Holdom:1982ew,Gelmini:1988sf,Coulson:1995nv} and have been raising increasing interest in the literature due to their connection with axion physics. 

Domain walls can then still play an important role in cosmology and their observational imprints may provide a new window into the physics of the early universe. Many recent studies have focused on the stochastic gravitational wave background generated by domain walls or on the potential production of primordial black holes (see e.g.~\cite{Hiramatsu:2013qaa, NANOGrav:2023hvm, Gelmini:2023ngs, Gouttenoire:2023ftk,Ferreira:2024eru, Dunsky:2024zdo, Gruber:2024jtc, Gruber:2024pqh}). However, the decay of domain walls is also expected to induce matter-density perturbations. These perturbations could give rise to non-linear structures potentially observable today. Although this domain-wall seeded structure formation has long been known to be subdominant, based on measurements of the anisotropies of the cosmic microwave background (which are consistent with perturbations having been seeded in the primordial universe), a quantification of their potential contribution is still missing. Here, we bridge this gap and analyze the potential role of domain wall networks in structure formation in more detail.\\

This paper is organized as follows. We briefly review the parameter-free model in Sec.~\ref{sec:Modelling}, where we also calculate the energies in the final stages of collapse and the rate of collapse events. Section \ref{sec:NonlinearStruc} focuses on the calculation of the resulting masses of the non-linear objects seeded by domain walls and the associated halo mass function. In Sec.~\ref{sec:Biased}, we explore the implications of our analysis for biased domain wall networks and examine the possibility that these networks may  potentially provide an explanation for an apparent mass excess at large redshifts suggested by recent James Webb Space Telescope (JWST) data. Finally, we present our conclusions in Sec.~\ref{sec:Conclusions}.%

Unless explicitly stated otherwise, we will work in natural units with $\hbar=c=1$ and use the cosmological parameters as measured by the Planck mission \cite{Planck}, such that the current density parameters are $\Omega_{\rm m0}=0.315$, $\Omega_{\rm r0}=9.1476\cdot10^{-5}$ and $\Omega_{\rm \Lambda 0}=1-\Omega_{\rm m0}-\Omega_{\rm r0}$ for matter, radiation and the cosmological constant, respectively, and the Hubble parameter at the present time is given by $H_0=67.4\ \rm{km\ s^{-1}\ Mpc^{-1}}$.

\section{Domain wall network: Modelling}
\label{sec:Modelling}
In this section, we provide an overview of the parameter-free velocity-dependent one-scale model \cite{parameterfree,Avelino:2020ubr} for the evolution of domain wall networks, upon which this work is based. We then use this model to calculate the energy available in the final stages of a single domain wall collapse, as well as the rate of such collapses within our horizon volume (which are the ones which could potentially generate observable signatures). The predictions of the parameter-free model have been shown to agree well with numerical simulations for fast enough expansion rates \cite{parameterfree,Avelino:2020ubr}. Throughout this paper, we will consider a flat $3+1$-dimensional Friedmann-Lema\^itre-Robertson-Walker universe with line element
\begin{equation}
    ds^2=a^2[\eta]\left(d\eta^2-d{\bf x}\cdot d{\bf x}\right),
    \label{eqn:FLRW}
\end{equation}%
where $a$ is the cosmological scale factor, $t$ is the physical time, $\eta=\int dt/a$ is the conformal time, and ${\bf x}$ are comoving coordinates. 
\subsection{The parameter-free model}
\label{sec:paramfree}
The parameter-free model introduced in \cite{parameterfree} allows for the description of the evolution of a domain wall network in terms of the successive collapse of increasingly larger domains based on a statistical approach free from adjustable parameters. It assumes that a network of infinitely thin spherical\footnote{The parameter-free model also considers cylindrical walls as a limiting case.} domain walls starts from rest at an initial conformal time $\eta_{\rm i}$ with a probability distribution of initial radii $q_{\rm i}$ given by \citep{parameterfree}
\begin{equation}
    P\left[q_{\rm i}\right]=3\eta_{\rm i}^3q_{\rm i}^{-4}\Theta\left[q_{\rm i}-\eta_{\rm i}\right],
    \label{eq:probs}
\end{equation}
where $\Theta$ denotes the Heaviside step function. This distribution ensures that the energy density $\rho$ of the network scales as $\rho\propto t^{-1} \propto (a\eta)^{-1}$, thus matching the scaling regime seen in numerical simulations. Individual walls detach from the Hubble flow and rapidly start to collapse as soon as they enter the Hubble horizon. These walls completely decay when their radius reaches $q=0$, such that at a conformal time $\eta$, all walls with an initial radius below a certain cutoff $q_{\rm i*}\left[\eta\right]$ cease to be part of the network. For simplicity, let us start by assuming that the evolution of the scale factor is well approximated by a power-law of the form $a\propto t^\lambda \propto \eta^{\lambda/(1-\lambda)}$ for $\lambda \in (0,1)$. According to the Nambu-Goto action as generalised to $2+1$ dimensions (also known as the Dirac action), the equations of motion for a spherical domain wall with radius $q$ are \citep{parameterfree}
\begin{align}\label{eqn:equationsofmotion}
    \dot{q}&=-v\\
    \dot{v}
    &=\gamma^{-2}\left(\frac{2}{q}-3\mathcal{H}v\right),
    \label{eqn:equationsofmotion2}
\end{align}
where dots denote derivatives with respect to the conformal time, $v$ is the velocity of an individual wall, $\gamma$ is the Lorentz factor defined as $\gamma=\left(1-v^2\right)^{-1/2}$ and $\mathcal{H}=\mathcal{H}[\eta]=\dot{a}/a$. Invariance of these governing equations under the transformations $q\rightarrow \alpha q$, $\eta \rightarrow \alpha \eta$ for $\alpha \in \mathbb{R}_{>0}$ allows for the description of a wall in terms of the rescaled radius $\tilde{q}\left[\tau\right]=q/q_{\rm i}$, the velocity $v\left[\tau\right]$ and $\gamma\left[\tau\right]$, where $\tau=\eta/q_{\rm i}$, such that $\tilde{q}\left[0\right]=1$~\citep{parameterfree}. The wall then decays at a time $\tau_*=\eta/q_{\rm i*}$, so that $\tilde{q}\left[\tau_*\right]=0$. Here and in the remainder of this work, the subscript $_*$ denotes quantities evaluated at the time of wall decay, $t_*$.
\begin{figure}[t]
\centering
\vspace{2pt}
\includegraphics[width=0.993\linewidth]{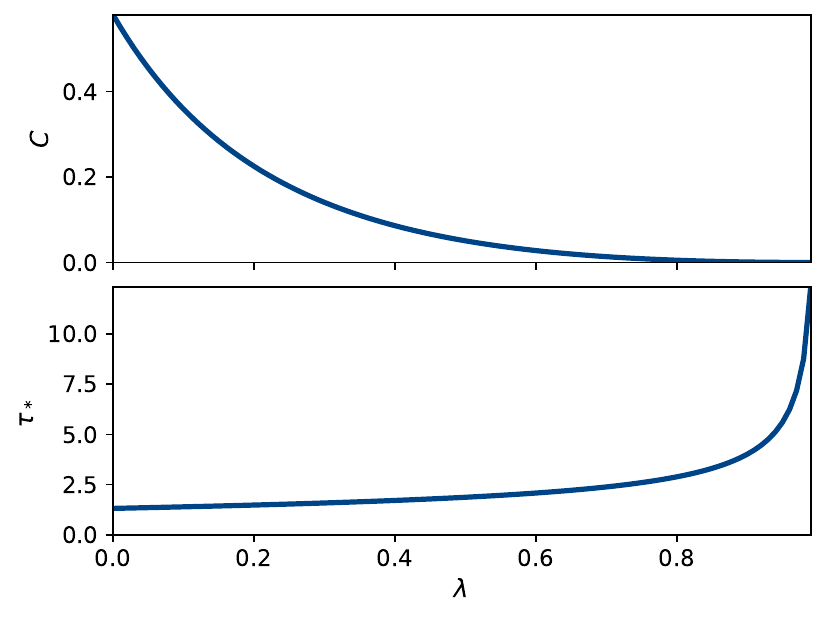}%
\caption{Dependence of the scale-invariant parameter $C$, defined in Eqs.~\eqref{eqn:decayenergies} and~\eqref{eqn:Cpar}, and the dimensionless collapse time $\tau_*$ on the expansion rate parameter $\lambda$. As $\lambda$ increases, Hubble damping becomes more efficient, leading to a longer wall collapse time.\label{fig:Candtaus}}
\end{figure}

\subsection{Decay energies}
\label{sec:Energies}
In the absence of friction due to non-minimal coupling with other fields, we can estimate the wall decay energy, $E_*$, of an individual collapse occurring at a redshift $z_*$ in the parameter-free model as \citep{parameterfree}
\begin{equation}
    E_*=4\pi \varepsilon \sigma_{\rm Zel} a_*^2\eta_*^2 C(\lambda)\,,
    \label{eqn:decayenergies}
\end{equation}
where 
\begin{equation}
    C(\lambda)=\frac{\gamma_* q_*^2}{\eta_*^2}
\label{eqn:Cpar}
\end{equation}%
is a scale-invariant quantity. The adimensional parameter $\varepsilon=\sigma_{\rm w}/\sigma_{\rm Zel}$ denotes the energy per unit area (or tension) of the wall, rescaled by the upper limit corresponding to the Zel'dovich bound. Solving the governing equations given in Eqs.~\eqref{eqn:equationsofmotion} and \eqref{eqn:equationsofmotion2}, we can determine the parameter $C$, as well as the decay time $\tau_*$ numerically, as shown in Fig.~\ref{fig:Candtaus}.\\

\begin{figure}[t]
\vspace{0pt}
\includegraphics[width=0.995\linewidth]{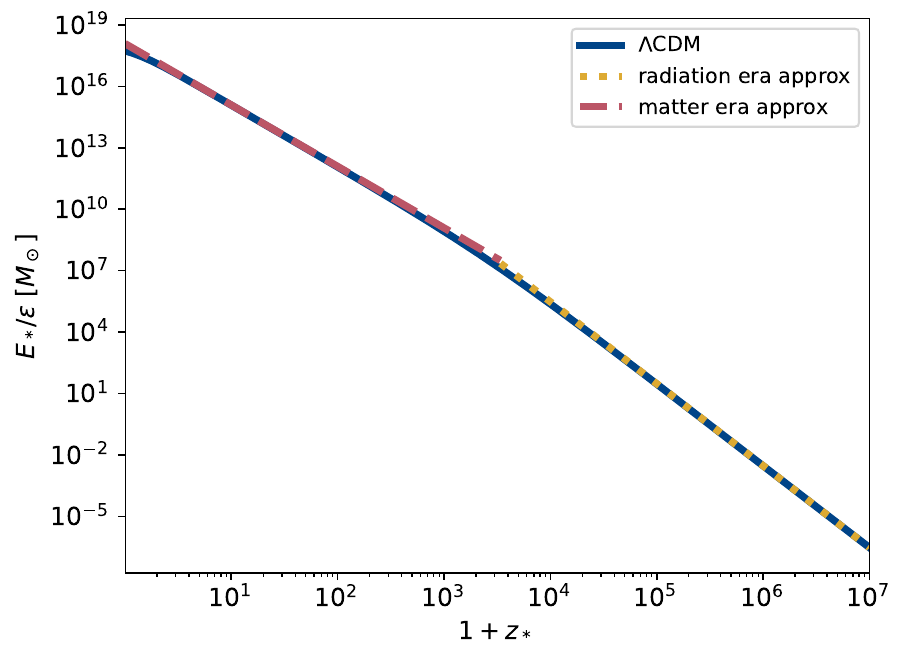}%
\caption{Decay energies of individual wall collapses and their dependence on the decay redshift, $z_*$, and rescaled wall tension $\varepsilon=\sigma_{\rm w}/\sigma_{\rm Zel}$, where $\sigma_{\rm Zel}$ denotes the upper limit derived from the Zel'dovich bound. This plot displays the full numerical results (solid blue line) alongside the analytical approximations for the matter (red  dashed line) and radiation eras (yellow dotted line) in Eqs.~\eqref{eqn:energyparameters} and~\eqref{eqn:energyparameters2}. The influence of the cosmological constant is visible at low redshift as a slight dampening of the expected available decay energy.\label{fig:energies}}
\end{figure}

Considering a full $\Lambda {\rm CDM}$ universe evolving according to 
\begin{equation}
    H^2=H_0^2 \left[\Omega_{\rm m0} a^{-3}+ \Omega_{\rm r0}a^{-4} +\Omega_{\rm \Lambda 0}\right],
\end{equation}
where $H$ is the Hubble parameter, one cannot, in general, assume that $a\propto t^\lambda$ (except deep in the radiation or matter eras). However, we may define an effective $\lambda_{\rm eff}$ as 
\begin{equation}
    \lambda_{\rm eff} (t)=\frac{d\log{a}}{d\log{t}},
    \label{eq:lambdaeff}
\end{equation}%
such that one can assume instantaneously that $a$ is a linear function of $t^{\lambda_{\rm eff}}$. We display the redshift dependence of $\lambda_{\rm eff}$ in Fig.~\ref{fig:lambdaeff}. It shows that $\lambda_{\rm eff} \sim 1/2$ and $\lambda_{\rm eff} \sim 2/3$ in the radiation- and matter-dominated eras, respectively, and that its value exhibits a sharp increase as the universe transitions into the $\Lambda$-dominated regime. The knowledge of the cosmological evolution of this parameter allows us to estimate the full decay energies $E_*$ associated to the collapse of individual domain walls throughout cosmic history in a $\Lambda$CDM background. The results are displayed in Fig.~\ref{fig:energies} (solid blue line).\\

One may also derive simple analytical approximations for the energy of collapsing walls during the matter and radiation eras. In \cite{AnalyticalScalingSols}, the following approximate solution to Eqs.~\eqref{eqn:equationsofmotion} and \eqref{eqn:equationsofmotion2} was derived\footnote{It should be noted that, although the accuracy of this approximation is higher in the non-relativistic regime, it may also be employed in the relativistic limit (see \cite{AnalyticalScalingSols} for more details).}:
\begin{figure}[t]
\centering
{\includegraphics[width=\linewidth]{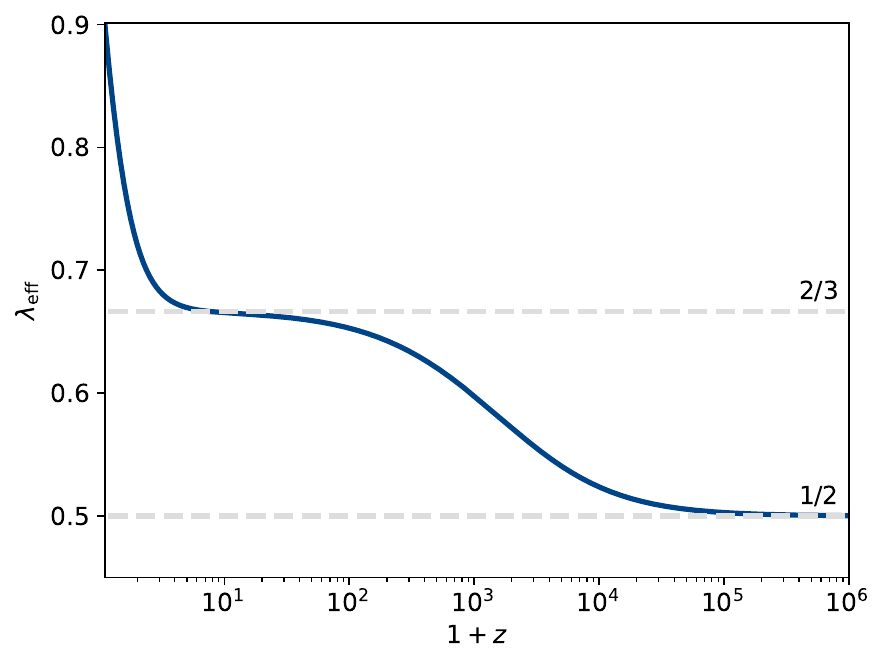}}%
\caption{Evolution of the effective parameter $\lambda_{\rm eff}$, characterizing the instantaneous rate of expansion in a $\Lambda$CDM background. It can be seen that this approaches $\lambda=2/3$ in a matter-dominated and $\lambda=1/2$ in a radiation-dominated regime, respectively. At low redshift, a sharp growth of $\lambda_{\rm eff}$ is triggered when the universe enters the $\Lambda$-dominated era.\label{fig:lambdaeff}}
\end{figure}%
\begin{align}
    &v^2\gamma \tilde{q}^2=\frac{\tau^2}{\tau_*^4}\\
    &\tilde{q}=\sqrt{1-\left(\frac{\tau}{\tau_*}\right)^2}\\
    &\tau_*=\sqrt{\frac{2\lambda+1}{2(1-\lambda)}}\,,
    \label{eqn:approximation}
\end{align}%
which yields $\tau_*=\sqrt{7/2}$ for a purely matter-dominated universe with $\lambda=2/3$ and $\tau_*=\sqrt{2}$ for a purely radiation-dominated universe with $\lambda=1/2$. Together with Eq.~\eqref{eqn:decayenergies}, this allows us to estimate the energy associated to a domain wall collapse during the matter-dominated era as
\begin{align}
    E_{*,\rm m}&=f_{\rm m}\frac{64\pi}{49}\varepsilon\sigma_{\rm{Zel}}H_0^{-2}\Omega_{\rm{m}0}^{-1}\left(1+z_*\right)^{-3}\label{eqn:energyparameters}\\
    &\sim 1.23 \times 10^{18}M_\odot \varepsilon(1+z_*)^{-3}\,,
    \nonumber
\end{align}%
where $M_\odot$ denotes the solar mass and $z_*$ is the redshift at which the wall decays. Here, the correction factor $f_{\rm m}=0.22$  was introduced to improve the fit to the numerical results. For a collapse occuring during the radiation-dominated era, this is similarly given by
\begin{align}
    E_{*,\rm r}&=f_{\rm r}\pi\varepsilon\sigma_{\rm Zel}H_0^{-2}\Omega_{\rm r0}^{-1}\left(1+z_*\right)^{-4} \label{eqn:energyparameters2}\\
    &\sim 2.98 \times 10^{21}M_\odot \varepsilon \left(1+z_*\right)^{-4}\,,
    \nonumber
\end{align}
with $f_{\rm r}=0.20$. In the derivation of these expressions, we used the fact that $v\sim1$ in the final stages of the collapse \cite{AnalyticalScalingSols}. We display these approximations in Fig.~\ref{fig:energies} (yellow dotted and red dashed lines corresponding to the radiation and matter eras, respectively). Therein, one may see that these approximations provide an excellent fit to the numerical results throughout most of cosmic history, except during the radiation-matter transition and the transition to $\Lambda$-domination.\\

\subsection{Collapse Rate}
\label{sec:collapses}
At a time $\eta$, all walls with an initial radius $q_{\rm i}$ below a certain cutoff $q_{\rm i*}[\eta]$ have already decayed. The distribution of initial radii as determined by Eq.~\eqref{eq:probs} thus allows us to estimate the number of collapses per unit volume from a conformal time $\eta$ until $\eta+d\eta$ as
\begin{equation}
    \frac{N}{V}=\int_{q_{\rm i *}(\eta)}^{q_{\rm i*}(\eta+d\eta)} nP[q_{\rm i}]dq_{\rm i}\,.
    \label{eqn:N_V}
\end{equation}
Here, $n$ denotes the number density of walls at the time $\eta$ in the absence of collapse, written as $n=n_{\rm i}(a/a_{\rm i})^{-3}$, where $n_{\rm i}$ labels the initial value at the time $\eta_{\rm i}$ and $a_{\rm i}=a(\eta_{\rm i})$. Based on the scaling introduced in Sec.~\ref{sec:paramfree}, the cutoffs can be expressed as
\begin{equation}
    q_{\rm i*}=\frac{\eta}{\tau_*}\,.
    \label{eqn:cutoff}
\end{equation}
Using that $n_{\rm i}=\beta/(12\pi\eta_{\rm i}^3)$ and $\beta\sim 1.15$~\cite{parameterfree, Martins_2016}, and combining Eqs. \eqref{eqn:N_V} and \eqref{eqn:cutoff}, we are able to write the distribution of collapse events within the current physical particle volume as
\begin{equation}
    \frac{dN}{dz_*}=-\frac{\beta}{3}\frac{\eta_0^3}{\eta^4}\tau_*(\eta)^3\frac{d\eta}{dz_*}.
    \label{eq:distrib}
\end{equation}
Numerically, we may obtain this distribution in a $\Lambda$CDM background by resorting to the instantaneous rate of expansion in Eq.~\eqref{eq:lambdaeff} and using the evolution of the collapse time $\tau_*$ as shown in Fig.~\ref{fig:Candtaus} (bottom). Note however that some care must be taken in the computation of $\eta(z)$ using $\lambda_{\rm eff}$: as $z\to 0$, $\lambda_{\rm eff}$ approaches unity very quickly and consequently $\eta$ also increases very fast. To avoid overcounting the number of collapses as a result of this, we use, in this computation, $\lambda=2/3$ (i.e. the matter-era value) for low redshift as well. Since the cosmological constant becomes significant only at low redshift, this gives us a good estimate of the collapse rate up until the current time.\\
\begin{figure}
\vspace{0pt}
\includegraphics[width=\linewidth]{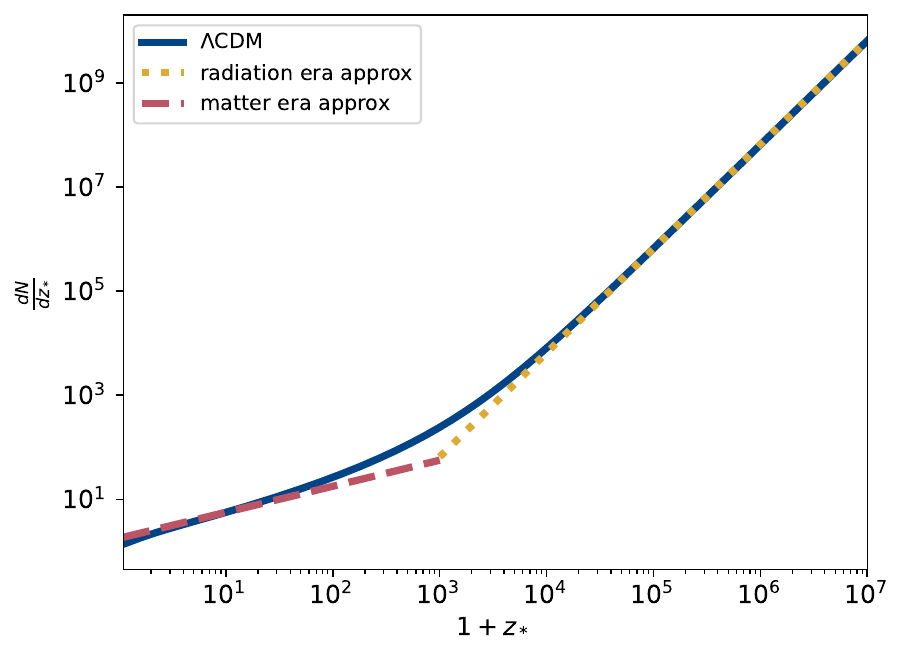}%
\caption{Distribution of domain wall collapse events within the current horizon volume. The numerical result for a full $\Lambda$CDM universe is shown as a solid blue line, together with analytical approximations for the matter (red dashed line) and radiation (yellow dotted line) eras as given in Eqs.~\eqref{eqn:distribution} and~\eqref{eqn:distribution2}, respectively. Since the network becomes more and more diluted as the universe expands, the number density of collapse events decreases steeply as the collapse redshift $z_*$ decreases as well.\label{fig:dist}}
\end{figure}

As before, we can derive approximations for this distribution  during the radiation and matter eras. In the matter-dominated era, taking into account that $\tau_*=(7/2)^{1/2}$, we have that
\begin{equation}
    \left(\frac{dN}{dz_*}\right)_{\rm m}=g_{\rm m}\frac{\beta}{6}\left(\frac{7}{2}\right)^\frac{3}{2}(1+z_*)^\frac{1}{2}\,,
    \label{eqn:distribution}
\end{equation}
where we have introduced the factor $g_{\rm m}=0.8$ to get a better fit to the full numerical results, which include the effect of the cosmological constant. To obtain an analytical approximation for the distribution of collapses in the radiation era, the effect of the radiation-matter transition needs to be taken into account. We model this transition as happening instantaneously at redshift $z_{\rm eq}=3390$ (as suggested by Planck data~\cite{Planck}) with the evolution of the scale factor being determined by $\lambda=1/2$ for $z>z_{\rm eq}$ and $\lambda=2/3$ for $z<z_{\rm eq}$. We then find that
\begin{equation}
     \left(\frac{dN}{dz_*}\right)_{\rm r}=g_{\rm r}\frac{\beta}{3}2^{\frac{3}{2}}\frac{\left(1+z_*\right)^2}{\left(1+z_{\rm eq}\right)^{\frac{3}{2}}}\,, 
     \label{eqn:distribution2}
\end{equation}
again including a fitting factor $g_{\rm r}=7.9$.\\

In Fig.~\ref{fig:dist}, we display the distribution of collapse events as a function of redshift, in a $\Lambda$CDM background, alongside the analytical approximations given in Eqs.~\eqref{eqn:distribution} and~\eqref{eqn:distribution2}. Therein, one can see that our analytical approximations generally provide a good description of the full numerical results, except in the range of redshifts corresponding to the radiation-matter transition. The number density of domain wall collapse events decreases steeply as $z_*$ decreases, since the domain wall network becomes less and less dense as the universe expands. We estimate that a total of about $10^6$ relevant collapse events --- i.e. collapses that are within our particle horizon --- may have happened since the radiation-matter transition and roughly $10^7$ since a redshift of $z=10^4$.

\section{Non-linear structures seeded by collapsing Domain Walls}
\label{sec:NonlinearStruc}
Domain wall collapses may act as seeds of density perturbations and result in the formation of non-linear objects. In this section, we characterise the masses of these domain-wall-seeded structures and determine the corresponding halo mass function. Throughout this paper, we model CDM and baryons as a single CDM fluid with current density parameter $\Omega_{\rm m0}=\Omega_{\rm CDM0}+\Omega_{\rm B0}$. This is justified by the fact that baryons constitute a subdominant fraction of the total matter density ($\Omega_{\rm B0}/\Omega_{\rm CDM0} \sim 0.2$) and, after recombination, effectively behave as CDM on mass scales $M \gtrsim 10^5 \, M_\odot$ \cite{Peebles1968}. This approximation simplifies the analysis without significantly affecting the accuracy of large-scale structure predictions, even for mass scales smaller than $10^5 \, M_\odot$ (since $\Omega_{\rm m 0} \sim \Omega_{\rm CDM 0}$).

\subsection{Matter-density perturbations}
\label{sec:perturb}
The collapse of a wall with the energy given in Sec.~\ref{sec:Energies} in a $\Lambda$CDM universe will cause a perturbation in the background matter-density field. We assume in the following that the associated comoving displacement is spherically symmetric around the centre of the collapsing wall and can thus be written as
\begin{equation}
    {\bf \Psi}=\Psi\left(t, {\bf q}\right)\hat{\bf e}_{\rm {\bf q}}\,,
    \label{eq:comov}
\end{equation}
where ${\bf q}$ are the comoving coordinates labelling the unperturbed positions of the CDM particles, $\hat{{\bf e}}_{\bf q}$ denotes a unit vector with origin at the centre of the wall and $q=|{\bf q}|$. We can then express the perturbed position of the CDM particles as
\begin{equation}
    {\bf r}=a\left[{\bf q}+{\bf \Psi}\left(t,{\bf q}\right)\right]\,.
    \label{eqn:perturbed}
\end{equation}%
Our aim is to characterise the number density and mass distribution of the non-linear objects created by these perturbations as they appear today. This means that we are interested in how the domain wall collapse affects the background CDM field within a radius given by the turnaround radius today, which is defined by $\dot{\bf r}(t_0, {\bf q})={\bf 0}$. We can therefore assume that the collapsing wall starts to significantly affect the surrounding matter as soon as it enters this surface. This corresponds to the finite-time presence of a point-mass-like gravitational source with the energy (or mass) $E_*$ that we derived in Sec.~\ref{sec:Energies}. In the linear regime, the evolution of the CDM density perturbations can be described using the Zel'dovich approximation~\cite{ZelApprox}: 
\begin{equation}
    \delta=-a\nabla\cdot {\bf \Psi},
\end{equation}
where ${\bf \Psi}$ evolves according to
\begin{equation}
    \ddot{{\bf \Psi}}(t,{\bf q})+2H\dot{{\bf \Psi}}(t,{\bf q})-4\pi G\overline{\rho}_{\rm{m}}{\bf \Psi}(t,{\bf q})=\frac{1}{a}{\bf a}_s\,.
    \label{eq:Psi}
\end{equation}
Here, $\overline{\rho}_{\rm m}$ denotes the average matter density, $G$ is the gravitational constant and ${\bf a}_s$ is the acceleration experienced by an arbitrary fluid element due to the collapsing domain wall. We will assume that the collapsing wall suddenly starts acting as a source for the perturbation at a time $t_\_=t_*-\Delta t$ and continues to do so until its decay at $t_*$, with $\Delta t=qa_*$ being the time taken for the wall to collapse. The radiation and scalar particles produced during the wall decay will continue to act as a source from the time $t_*$ until $t_+=t_*+\Delta t$, when they are able to escape the turnaround volume. We may then express ${\bf a}_{\rm{s}}$ as a gravitational source term acting from a time $t_\_$ until a time $t_+$ according to
\begin{equation}
    {\bf a}_{\rm{s}}=-\frac{GE_*}{a^2q^3}{\bf q}\Theta\left(t_+-t\right)\Theta\left(t-t_\_\right).
\end{equation}
For the full $\Lambda {\rm CDM}$ case, our definition of $\lambda_{\rm eff}$ as given in Eq.~\eqref{eq:lambdaeff} allows us to model a gradual transition between all the regimes, and to thus track the evolution of the generated perturbation throughout cosmic history.\\

One may also use Eq.~\eqref{eq:Psi} to obtain analytical approximations for the evolution of the comoving displacement generated by walls decaying in the matter and radiation eras, by again assuming that the universe is initially purely radiation dominated for $t<t_{\rm eq}$ and that it undergoes an instantaneous transition to matter domination at $t_{\rm eq}$. These yield, respectively:
\begin{widetext}
\begin{equation}
    {\bf \Psi}_{\rm m}\left(t,{\bf q}\right) = \begin{cases} 
          \frac{3}{2}\frac{GE_*}{q^3}{\bf q}t_i^2\left\{1-\frac{3}{5}\left(\frac{t}{t_*}\right)^\frac{2}{3}\left[1+\frac{2}{3}\frac{\Delta t}{t_*}\right]-\frac{2}{5}\left(\frac{t_*-\Delta t}{t}\right)\right\}\,, & t_*-\Delta t \leq t\leq t_*+\Delta t \\
          \frac{6}{5}\frac{GE_*}{q^3}{\bf q}t_i^2 \left\{\frac{\Delta t}{t}-\left(\frac{t}{t_*}\right)^{\frac{2}{3}}\frac{\Delta t}{t_*}\right\}\,, & t\geq t_*+\Delta t,
       \end{cases}
    \label{eq:comovana}
\end{equation}
\begin{equation}
    {\bf \Psi}_{\rm r}(t,{\bf q})=\begin{cases} 
     -\frac{GE_*}{q^3}{\bf q}t_i^{\frac{3}{2}}\left\{-2t_*^\frac{1}{2}\left[1-\frac{1}{2}\frac{\Delta t}{t_*}\right]\ln{\left(\frac{t}{t_*-\Delta t}\right)}+4\left(t^\frac{1}{2}-t_*^\frac{1}{2}\left[1-\frac{1}{2}\frac{\Delta t}{t_*}\right]\right)\right\}\,,& t_*-\Delta t \leq t\leq t_*+\Delta t\\
     -2\frac{GE_*}{q^3}{\bf q}t_i^{\frac{3}{2}}t_*^\frac{1}{2}\Delta t\ln{\left(\frac{t}{t_*}\right)}\,,& t_*+\Delta t \leq t\leq t_{\rm eq}\\
     -\frac{GE_*}{q^2}{\bf q}t_i^\frac{3}{2}t_*^{-\frac{1}{2}}\Delta t\left\{\frac{4}{5}\left[\frac{3}{2}-\ln{\left(\frac{t_{\rm eq}}{t_*}\right)}\right]\left(\frac{t_{\rm eq}}{t}\right)-\frac{6}{5}\left[1+\ln{\left(\frac{t_{\rm eq}}{t_*}\right)}\right]\left(\frac{t}{t_{\rm eq}}\right)^\frac{2}{3}\right\}\,,
     & t\geq t_{\rm eq}.
    \end{cases}
\end{equation}
\end{widetext}%
At low redshift, the increasing influence of the cosmological constant slows down the growth of any generated perturbations. This effect can be approximated using a growth factor $\zeta$  defined as \cite{CarrolPressTurner}  $\zeta(\Omega_{\rm m0}, \Omega_{\rm \Lambda 0}) = {\bf \Psi}_{\rm m+\Lambda}(t_0,{\bf q})/{{\bf \Psi}_{\rm m}(t_0, {\bf q})}$, where ${\bf \Psi}_{\rm m+\Lambda}={\bf \Psi}_{[\Omega_{\rm m0}, \Omega_{\Lambda 0}]}$ and ${\bf \Psi}_{\rm m}={\bf \Psi}_{[1,0]}$ are the perturbations measured today assuming they evolved in a flat universe with matter and a cosmological constant, and with matter only, respectively. \citet{Lahav} give an approximation to the rate of growth of these perturbations, which allows us in a similar way to define a factor $\theta$ such that $\theta(\Omega_{\rm m0}, \Omega_{\rm \Lambda 0}) = \dot{{\bf {\Psi}}}_{\rm m+\Lambda}(t_0,{\bf q})/\dot{{{\bf \Psi}}}_{\rm m}(t_0, {\bf q})$, where analogously $\dot{{\bf \Psi}}_{\rm m+\Lambda}=\dot{{\bf \Psi}}_{[\Omega_{\rm m0}, \Omega_{\Lambda 0}]}$ and $\dot{{\bf \Psi}}_{\rm m}=\dot{{\bf \Psi}}_{[1,0]}$. It was shown in~\cite{Avelino_1999} that these approximations hold well for the growth of linear perturbations in defect models. Throughout the next sections, we will use these correction factors to adjust our analytical results to account for the influence of the cosmological constant. For the cosmological parameters used here, we find $\zeta=0.787$ and $\theta=0.225$.

\subsection{Accreted masses}
\label{sec:masses}
\begin{figure}[t]%
\vspace{0.395pt}%
\includegraphics[width=0.99\linewidth]{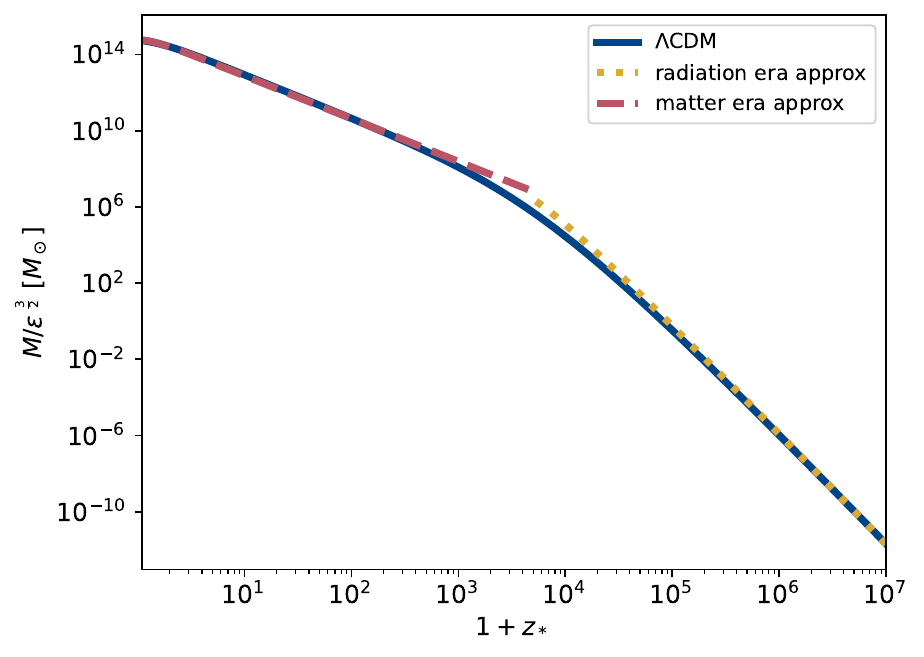}%
\vspace{3.15pt}
\caption{Present-day masses of non-linear objects seeded by the collapse of standard domain walls at a redshift $z_*$ in units of solar mass $M_\odot$ ($\varepsilon=\sigma_{\rm w}/\sigma_{\rm Zel}$ is the wall tension, rescaled by the upper limit derived from the Zel'dovich bound). Our numerical result for a $\Lambda$CDM universe is shown as a solid blue line, together with the approximations for the matter (dashed red line) and radiation (yellow dotted line) eras given in Eqs.~\eqref{eq:massmat} and~\eqref{eqn:massrad}, respectively. The mass of these objects can reach up to $\sim 10^{15} M_\odot$, with the most massive ones being created by walls collapsing recently in cosmic history, when the collapse rate is lower.}\label{fig:masses}
\end{figure}

We will now estimate the current masses of the non-linear objects seeded by domain walls as observed today. Eq.~\eqref{eqn:perturbed} allows us to write the turnaround condition ($\dot{{\bf r}}(t_0, {\bf q})={\bf 0}$) as
\begin{equation}
    {\bf q}_{\rm turn}=-\frac{1}{H_0}{\bf \dot{\Psi}}(t_0, {\bf q}_{\rm turn})-{\bf \Psi}(t_0, {\bf q}_{\rm turn}).
\end{equation}
Solving this for the turnaround radius $q_{\rm turn}=|\bf q_{\rm turn}|$, we can compute the current total mass $M$ of the collapsed object by assuming that it comprises all the matter initially enclosed within this comoving radius:\newpage
\begin{equation}
    M=\frac{4}{3}\pi a^3q_{\rm turn}^3\rho_{\rm m}\,.
    \label{eq:mass}
\end{equation}
In a realistic $\Lambda$CDM background, this computation may be performed numerically to obtain the masses of the non-linear objects seeded by domain wall collapses throughout cosmic history. The result is displayed in Fig.~\ref{fig:masses} (solid blue line). Therein, one may see that, depending on the decay redshift of the individual wall, the generated masses can reach up to $10^{15} M_\odot$, which corresponds roughly to the mass of a present-day galaxy cluster. The largest objects are, in fact, generated by walls collapsing in recent cosmological history, when the collapse rate is lower.\\

In our analytical approximation for a purely matter-dominated universe, this reduces to
\begin{eqnarray}
\label{eq:massmat}
 M_{\rm m}
    &=& \frac{4}{3}\left(\frac{6}{25}\right)^\frac{3}{4}\pi^\frac{1}{4}G^\frac{3}{4}\rho_{\rm m0}^{\frac{1}{4}}\left(E_{*,{\rm m}}\right)^\frac{3}{2} \times\\
    &\times& \left[h_{\rm m}\left(1+z_*\right)^{-1}+\tilde{h}_{\rm m}\left(1+z_*\right)^\frac{3}{2}\right]^\frac{3}{2}\nonumber\\ 
  &\sim& 10^{16}M_\odot \varepsilon^\frac{3}{2} \left[h_{\rm m}\left(1+z_*\right)^{-4}+\tilde{h}_{\rm m}\left(1+z_*\right)^{-\frac{3}{2}}\right]^\frac{3}{2}\,,\nonumber
\end{eqnarray}
for a non-linear object seeded by a wall collapsing at a redshift $z=z_*$, and where $E_*$ denotes the decay energy calculated in Sec.~\ref{sec:Energies}. We have corrected our approximation for the impact of the cosmological constant using the growth factors $h^{\rm m}=3\theta/2-\zeta$ and $\tilde{h}^{\rm m}=\theta+\zeta$. To derive an analytical approximation for the masses of non-linear objects seeded by wall collapses in the radiation era, we assume an abrupt transition between the radiation and matter regimes at $\tilde{z}_{\rm eq}$. This yields
\begin{widetext}
\begin{eqnarray}
        M_{\rm r} &=& \frac{\left(E_{*,{\rm r}}\right)^\frac{3}{2}}{9\sqrt{2}\Omega_{\rm m0}^{-1} \Omega_{\rm r0}^{{\frac{3}{4}}}} (H_0G)^\frac{1}{2} \rho_{\rm m0}^\frac{1}{4}\left[\frac{12}{5}\left(1+2\ln{\left\{\frac{1+z_*}{1+\tilde{z}_{\rm eq}}\right\}}\right)\left(1+\tilde{z}_{\rm eq}\right)^{-1}+\left(\frac{3}{2}-2\ln{\left\{\frac{1+z_*}{1+\tilde{z}_{\rm eq}}\right\}}\right)\left(1+\tilde{z}_{\rm eq}\right)^{\frac{3}{2}}\right]^\frac{3}{2}\label{eqn:massrad}\\
&\sim&  10^{14}M_\odot\varepsilon^\frac{3}{2}\left[\frac{12}{5}\left(1+2\ln{\left\{\frac{1+z_*}{1+\tilde{z}_{\rm eq}}\right\}}\right)\left(1+\tilde{z}_{\rm eq}\right)^{-5}+\left(\frac{3}{2}-2\ln{\left\{\frac{1+z_*}{1+\tilde{z}_{\rm eq}}\right\}}\right)\left(1+\tilde{z}_{\rm eq}\right)^{-\frac{5}{2}}\right]^\frac{3}{2}\,,\nonumber
\end{eqnarray}
\end{widetext}%
where we have used a factor $\tilde{z}_{\rm eq}/z_{\rm eq}=1.33$ to fit to our numerical results. We also display these approximations in Fig.~\ref{fig:masses}, where one may see that they provide an excellent description of the full numerical result. 
\subsection{Halo Mass Function}
\label{sec:halomass}
Here, we estimate the number of non-linear objects generated by the collapse of individual domain walls. This will be characterised by the halo mass function, defined as the number $n$ of collapsed objects per unit volume, per unit mass. Based on our estimation of the distribution of collapse events derived in Sec.~\ref{sec:collapses}, and our calculation of the mass dependence on the collapse redshift in Sec.~\ref{sec:masses}, it is possible to compute the halo
\begin{figure}[H]%
\vspace{-0pt}
\centering
\includegraphics[width=0.99\linewidth]{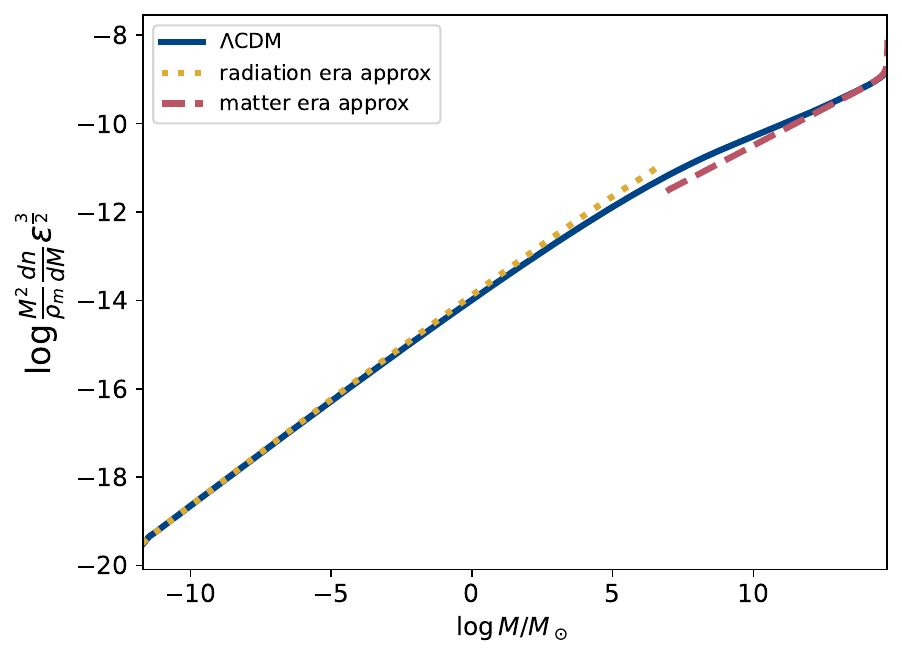}%
\vspace{-3.2pt}
\caption{Halo mass function for non-linear objects seeded by the collapse of standard domain walls in a $\Lambda$CDM universe. A comparison with the observed halo mass function as presented in \cite{HaloMass} reveals that the potential contribution of standard domain wall collapses to structure formation increases significantly with increasing mass, but that domain wall seeded objects can only make up a fraction of $\sim 10^{-9}$ of objects with mass of the order of $10^{10}M_\odot$ and $\sim 10^{-6}$ of objects with masses of $\sim10^{15}M_\odot$. Collapses of standard domain walls therefore make up a clearly subdominant contribution to overall structure formation.\label{fig:HaloMasses}}
\end{figure}%
\hspace{-10pt}mass function for the non-linear objects seeded by domain wall collapses. Our full numerical results for a $\Lambda$CDM universe are shown in Fig.~\ref{fig:HaloMasses}. Comparing with the values for the observed halo mass function given by \citet{HaloMass}, we can estimate the contribution from objects seeded by domain walls as making up, at most, a fraction of $\sim 10^{-9}$ of objects with masses of the order of $10^{10}M_\odot$ and $\sim 10^{-6}$ of objects with masses of the order of $10^{15}M_\odot$ (approximately the characteristic mass of galaxy clusters today). The collapse of standard domain walls could thus provide, at most, a subdominant contribution to structure formation.\\

Using Eqs.~\eqref{eqn:distribution},~\eqref{eqn:distribution2},~\eqref{eq:massmat} and~\eqref{eqn:massrad}, we may again derive analytical approximations to the halo mass function associated with domain walls collapsing either during the matter or the radiation eras within our particle horizon volume $V_{\rm p}$:
\begin{widetext}\vspace{5pt}
\begin{equation}
    \left(V_{\rm p}\frac{dn}{dM}\right)_{\rm m}\sim \frac{\beta}{9}\left(\frac{7}{2}\right)^{\frac{3}{2}}\left(10^{16}M_\odot\right)^{-1}\frac{\varepsilon^{-\frac{3}{2}}g_{\rm m}(1+z_*)^{\frac{1}{2}}}{\left[h_{\rm m}(1+z_*)^{-4}+\tilde{h}_{\rm m}(1+z_*)^{-\frac{3}{2}}\right]^\frac{1}{2}\left[-4h_{\rm m}(1+z_*)^{-5}-\frac{3}{2}\tilde{h}_{\rm m}(1+z_*)^{-\frac{5}{2}}\right]}\,,
\end{equation}
\begin{equation}
        \left(V_{\rm p}\frac{dn}{dM}\right)_{\rm r}\sim\frac{\beta}{9}2^\frac{3}{2}\left(10^{14}M_\odot\right)^{-1}\frac{\varepsilon^{-\frac{3}{2}}(1+z_*)^3\left(1+\tilde{z}_{\rm eq}\right)\left(\frac{12}{5}(1+\tilde{z}_{\rm eq})^{-\frac{5}{2}}-1\right)^{-1}}{\left[\frac{12}{5}\left(1+2\log{\{\frac{1+z_*}{1+\tilde{z}_{\rm eq}}\}}\right)(1+\tilde{z}_{\rm eq})^{-2}+\left(\frac{3}{2}-2\log{\{\frac{1+z_*}{1+\tilde{z}_{\rm eq}}\}}\right)(1+\tilde{z}_{\rm eq})^{\frac{1}{2}}\right]^{\frac{1}{2}}}\,.
\end{equation}
\end{widetext}%
Note that the dependence of these halo mass functions on the mass $M$ is implicit, but becomes apparent when considering the mass dependence on $z_*$ given in Eqs.~\eqref{eq:massmat} and~\eqref{eqn:massrad}. We display these together with our full numerical results in Fig.~\ref{fig:HaloMasses} (dotted yellow and dashed red lines for the radiation and matter era approximations, respectively).

\section{Biased networks and observational imprints}
\label{sec:Biased}
Until now, we have considered standard stable networks, in which walls of increasing size successively collapse after entering the Hubble horizon. These networks are assumed to persist until the current day and are therefore tightly constrained by CMB anisotropy measurements, meaning that the density perturbations associated to the collapse of such walls can only minimally contribute to the non-linear structures observed today. However, if the domain wall network is unstable and fully decays before the present time, the stringent constraints that result from the Zel'dovich bound may be evaded (especially if this decay happens before photon decoupling). Unstable domain wall networks, also known as biased domain wall networks, arise naturally if one of the vacua of the model is favoured, either by initial conditions (population bias) or because there is an asymmetry between the vacua (potential bias)~\cite{Gelmini:1988sf,Coulson:1995nv,Larsson:1996sp}. As a matter of fact, the latter scenario may arise in several relevant particle physics scenarios~\cite{Holdom:1982ew,Sikivie:1982qv,Gelmini:1988sf,Abel:1995wk,Riva:2010jm,Hiramatsu:2012sc,Gelmini:2020bqg,Perivolaropoulos:2022txg}. Although, in this case, there is an explicit symmetry breaking --- as there is an energy difference $\delta_{\rm V}$ between the minima of the model --- these networks may be relatively long lived provided that $\delta_{\rm V}$ is small. The volume pressure caused by this energy difference (that gives domains with the lowest energy density a tendency to expand) has to overcome the surface pressure caused by domain wall tension to be effective, something that only happens once the typical size of the domains reaches $R_{\rm bias} \sim \sigma/\delta_{\rm V}$~\cite{Coulson:1995nv,Larsson:1996sp,Avelino:2008qy}. For smaller domains, the impact of this volume pressure on wall dynamics is negligible\footnote{Note that in models with  a non minimal coupling to matter, other effects may arise that could impact the evolution of the network~\cite{Perivolaropoulos:2022txg, Christiansen:2023tfy}.}. Numerical simulations~\cite{Coulson:1995nv,Larsson:1996sp,Correia:2014kqa,Correia:2018tty,Cyr:2025nzf,Notari:2025kqq} show that biased domain wall networks, in fact, evolve as standard domain walls, in a linear scaling regime, until the typical domain size reaches $R_{\rm bias}$. Once this happens, this volume pressure becomes important, causing the domains of true vacuum to expand and triggering a fast complete decay of the domain wall networks. Although the physical mechanism driving network decay differs, the overall evolution of wall networks with a population bias is qualitatively very similar to that of networks with a potential bias~\cite{Coulson:1995nv,Hindmarsh:1996xv,Larsson:1996sp,Correia:2014kqa,Correia:2018tty}.\\

In this section, we will consider the impact of biased domain wall networks on the production of non-linear objects. We will assume, as suggested by numerical simulations, that these walls evolve in the typical linear scaling regime (which is well described by the parameter-free model) up to a redshift $z_{\rm bias}$ at which bias becomes important and the final decay of the network occurs. Since these walls behave as standard walls until $z_{\rm bias}$, the evolution of the network is still determined by the individual collapse of increasingly large domains that detach from the Hubble flow. These individual collapses may, as for standard walls, seed perturbations that generate non-linear objects at smaller redshifts. In particular, this means that our analysis in Secs. \ref{sec:Modelling}-\ref{sec:NonlinearStruc} remains applicable. However, in the case of biased walls, perturbations would only be seeded for $z_* \geq z_{\rm bias}$ (where, as before, $z_*$ is the redshift in which these individual collapses occur), because the network disappears at $z_{\rm bias}\gg 1$. This allows for higher wall tensions (such that the tension parameter $\varepsilon$ can now be larger than unity), and thus higher seed energies, which result in larger density perturbations. Note however that these walls must never dominate the cosmic energy budget, which results in a conservative bound on the allowed tensions~\cite{Gruber:2024jtc} given by
\begin{equation}
    \sigma_{\rm max}=\frac{3H_{\rm bias}\lambda_{\rm bias}\xi_{\rm bias}}{8\pi G}.
    \label{eqn:maxten}
\end{equation}
Here, the subscript `${\rm bias}$' indicates that the corresponding quantities should be evaluated at the time of the final decay of the network ($\lambda_{\rm bias}=1/2$ and $\xi_{\rm bias}=1.532$ for $z_{\rm bias}$ during the radiation era, and $\lambda_{\rm bias}=2/3$ and $\xi_{\rm bias}=1.625$ during the matter era). While this broadens the range of observationally permitted tensions --- especially at higher $z_{\rm bias}$ --- the framework developed for standard walls remains applicable.\\

\begin{figure}
\vspace{0pt}
    \centering
\includegraphics[width=\linewidth]{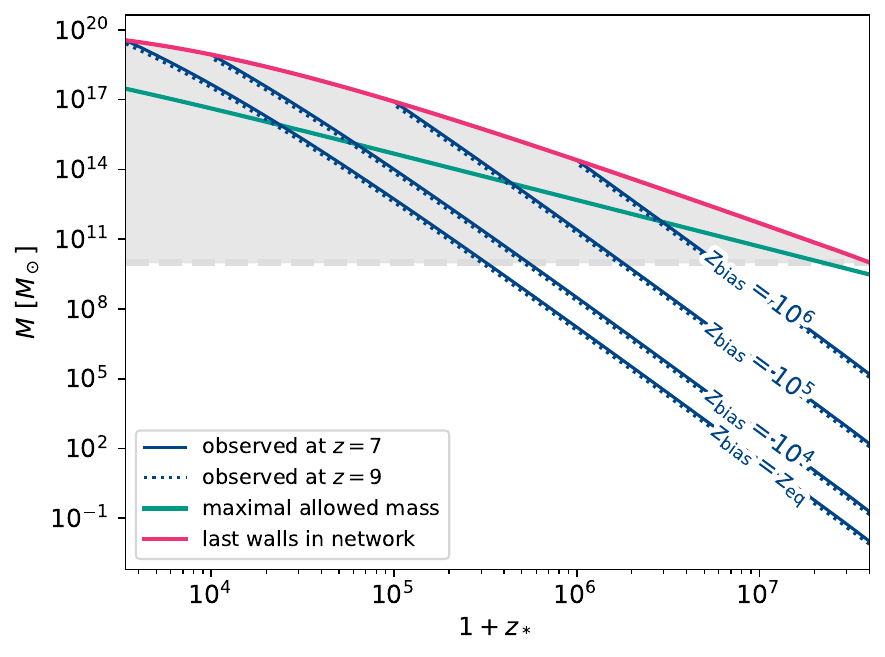}
    \caption{Maximum masses of non-linear objects seeded throughout the evolution of biased networks --- computed using the maximum allowed tension defined in Eq.~\eqref{eqn:maxten} ---  for a range of network decay redshifts $z_{\rm bias}$. Our results are presented for structures observed at $z=7$ (solid blue lines) and $z=9$ (dotted blue lines). The grey shaded region highlights the  parameter space where such networks can provide a contribution to objects with masses $M\geq 10^{10}M_{\odot}$, potentially alleviating observed tensions. The pink line shows the maximum masses reached at each $z_{\rm bias}$ (seeded by the last walls collapsing before $z_{\rm bias}$). The green line indicates the limit of applicability of our model.}
    \label{fig:halomassbiased}
\end{figure}

Compellingly, we found that these biased networks can generate a significant number of non-linear structures with masses $M\geq 10^{10} M_\odot$, potentially accounting for the observed excess of massive galaxies at redshifts of $z\gtrsim 7$ indicated by recent JWST observations (e.g.~\cite{Labb__2023, Boylan_Kolchin_2023}). Figure~\ref{fig:halomassbiased} shows the maximum mass --- computed using the maximum allowed wall tension for a particular $z_{\rm bias}$ as given in Eq.~\eqref{eqn:maxten} --- of non-linear structures observable at $z=7$ and $z=9$, seeded by networks decaying at various values of $z_{\rm bias}$ (blue solid and dotted lines, respectively). Note that the generated masses scale with $\varepsilon^{3/2}$ (see Eqs.~\eqref{eq:mass}-\eqref{eqn:massrad}), such that, as one lowers $z_{\rm bias}$ (and the maximum tension as a result), the maximal allowed mass also decreases. The results shown in Fig.~\ref{fig:halomassbiased} thus correspond to vertical translations of those presented in Fig.~\ref{fig:masses}, normalised to be observed at a redshift $z=7$ (solid blue line) and $z=9$ (dotted blue line) and cut at $z_*=z_{\rm bias}$. The grey shaded region highlights the mass range above $10^{10}M_\odot$, where biased domain walls could potentially help to alleviate current observational tensions. The solid pink line represents our estimate of the maximum mass of the non-linear objects seeded by domain wall networks decaying at each $z_{\rm bias}$. These are the objects associated to domain walls collapsing immediately before the network disappears (when $z_*=z_{\rm bias}$). The solid green line indicates the boundary beyond which our model's assumptions --- such as independent collapse --- begin to break down. This limit is obtained by requiring that the mass of each non-linear object does not exceed the total mass enclosed within the horizon volume at the time of the collapse of the corresponding domain wall. This figure clearly shows that biased networks can seed non-linear structures with masses far exceeding those permitted for standard walls  at early cosmological times (see Fig.~\ref{fig:masses}) and may therefore potentially play a relevant role in the formation of early massive structures.\\

\begin{figure}
    \centering
    \includegraphics[width=\linewidth]{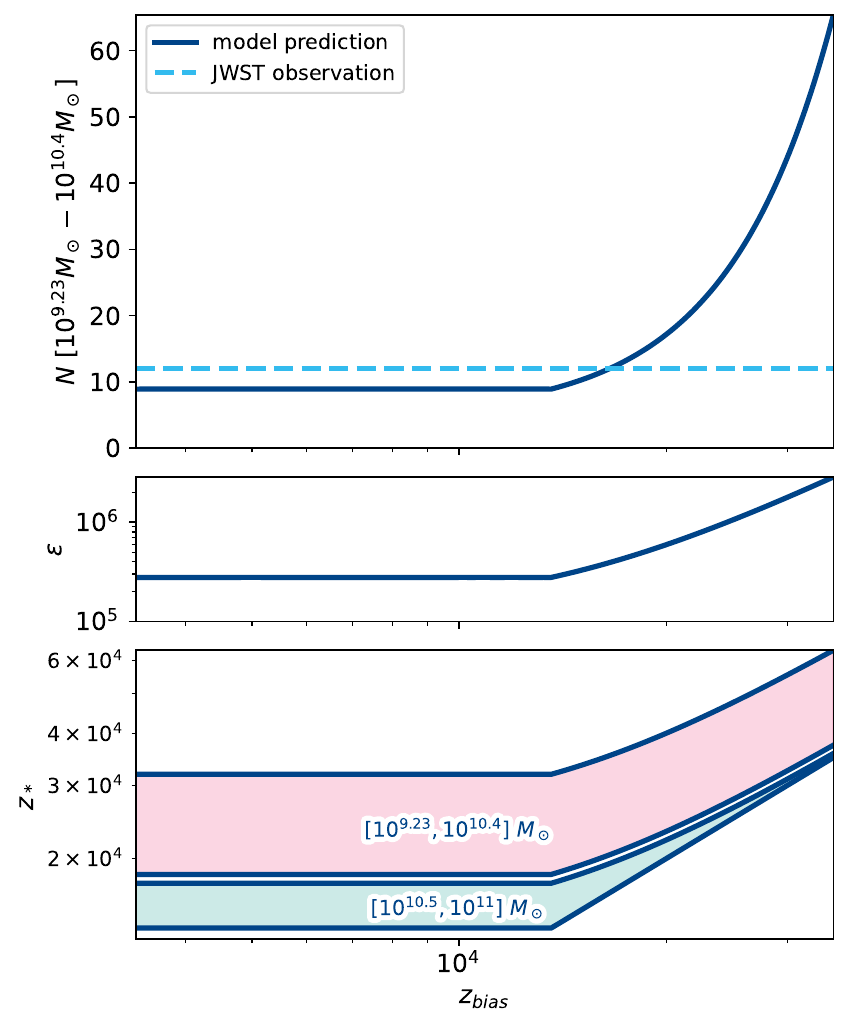}
    \caption{Region of parameter space in which our model may provide an explanation for the mass excess observed at JWST. Top panel: Predicted number of objects within the $[10^{9.23},10^{10.4}]M_\odot$ mass range
    based on our model, when normalised to a single object in the $[10^{10.5}, 10^{11}]M_\odot$ range (solid dark blue line), as a function of network decay redshift $z_{\rm bias}$. Observed values as given by \cite{Labb__2023} are shown as a dashed light blue line. Middle panel: Tension of the biased domain wall network that would generate a single object in the $[10^{10.5},10^{11}] M_\odot$ range in the observed area, expressed in terms of the adimensional parameter $\varepsilon=\sigma_{\rm w}/\sigma_{\rm Zel}$ (where $\sigma_{\rm Zel}$ is the tension corresponding to the Zel'dovich bound), as a function of $z_{\rm bias}$ (solid dark blue line). Bottom panel: range of decay redshifts of the walls that may seed objects in the mass ranges $[10^{9.23},10^{10.4}] M_\odot$ (pink shaded area) and $[10^{10.5},10^{11}] M_\odot$ (green shaded area) as a function of $z_{\rm bias}$. Our model appears to be in reasonable agreement with the observed counts up to $z_{\rm bias}=2.5\times 10^4$. Beyond that, the wall tension has to increase to enable the generation of an object in the $[10^{10.5},10^{11}] M_\odot$ mass range. In addition, the disappearance of the network means that the production of the remaining objects is pushed to higher redshift, causing a sharp increase in the number of objects in the mass range observed by JWST.}
    \label{fig:compJWST}
\end{figure}

\citet{Labb__2023} report, in particular, one object with a mass of $10^{10.89}M_\odot$, along with twelve others in the range of $10^{9.23}M_\odot$ to $10^{10.4}M_\odot$, within a survey field of $38\ {\rm arcmin^2}$. The existence of such massive objects at these redshifts is unexpected in standard structure formation models, which motivates us to investigate whether biased domain wall networks may provide an explanation for these results. Here, we will assume that the observed region is typical (and discuss the implications of this assumption in Sec.~\ref{sec:Conclusions}). Normalising to a single high-mass object in the $[10^{10.5},10^{11}]M_\odot$ bin --- i.e., by finding the tension that, for each $z_{\rm bias}$, corresponds to the production of one single object with a mass in this range --- our model allows us to predict the number of objects in other mass bins as a function of $z_{\rm bias}$. The top panel of Fig.~\ref{fig:compJWST} shows the predicted number counts in the $[10^{9.23},10^{10.4}] M_\odot$ mass range. To calculate these we used Eq.~\eqref{eq:mass} (to calculate the masses) and Eq.~\eqref{eq:distrib} (to determine the number of wall collapses at each redshift). Our model predicts approximately nine objects in the observed range for $z_{\rm bias} \lesssim 1.4 \times 10^4$. Beyond this network decay redshift, objects with masses in the range $[10^{10.5},10^{11}]M_\odot$ must be generated by walls collapsing in the very last stages of evolution of the biased network, at redshifts $z_* \sim z_{\rm bias}$, since collapses at higher redshift seed less massive objects (see Fig.~\ref{fig:masses}). As shown in Fig.~\ref{fig:compJWST} (middle panel), this implies that increasingly larger values of the wall tension will be required in order to reach sufficient mass as $z_{\rm bias}$ increases. A larger value of $z_{\rm bias}$ also implies that the collapse redshifts $z_*$ of the walls seeding these structures are shifted to larger values (see Fig.~\ref{fig:compJWST}, bottom panel), resulting in the sharp increase in the predicted number of objects seen in the top panel of this figure (consistent with the distribution of collapse events shown in Fig.~\ref{fig:dist}).\\

Our model therefore successfully reproduces the number of observed objects in the $[10^{9.23},10^{10.4}] M_\odot$ bin for a broad range of $z_{\rm bias}$, up to $z_{\rm bias}\sim 2.5 \times 10^4$, for a tension of $\varepsilon \sim 2.8\times 10^5$, or $\sigma_{\rm w}\sim (65\ {\rm MeV})^3$. If the observed excess is attributed to biased domain walls, this constrains the network decay redshift roughly to the interval $z_{\rm bias} \in [z_{\rm eq}, 2.5\times 10^4]$. We impose this lower cutoff, corresponding to the redshift of matter-radiation equality, since collapses during the matter era, with these substantial values of tension, should leave visible signatures in the CMB. More realistic constraints on the tension based on CMB measurements --- which have currently only been derived for standard walls surviving until the present day (see~\cite{Sousa:2015cqa}) --- could potentially reduce this redshift interval. As a matter of fact, a straightforward extrapolation of the Zel'dovich bound to biased walls --- assuming that walls can make up at most a fraction of $10^{-5}$ of the cosmic energy budget ---, would reduce the allowed network decay redshifts to a narrow range around $z_{\rm bias}\sim 10^4$. However, a detailed analysis of the CMB signatures of the scenarios discussed here would be necessary before the viability of the model as an explanation for the mass excess in JWST data could be definitely established.\\

In addition to the mass range observed by JWST, our model also predicts the existence of a small number of objects with masses exceeding $10^{11} M_\odot$, depending on $z_{\rm bias}$. For a tension $\varepsilon \sim 2.8\times 10^5$ (for which our model reproduces the results given in \cite{Labb__2023}), we predict on average one object in the mass range  $[10^{11},10^{11.5}] M_\odot$ within a survey field twice the size of the one observed, for $z_{\rm bias}<1.3\times 10^4$. Similarly, one object in the $[10^{11.5},10^{12}] M_\odot$ range is predicted within a survey field four times larger than the one observed, for $z_{\rm bias}<9.5\times 10^3$. Future observations, in particular if such high-mass objects are detected, would allow us to significantly constrain our model. At the moment, the lack of such detections favours an earlier decay of the network.

\section{Discussion and Conclusions}
\label{sec:Conclusions}
We have shown that, while standard domain walls only subdominantly contribute to structure formation, biased wall networks that have fully decayed by a redshift $z_{\rm bias}\gg 1$ can have a significantly greater impact. In particular, our results show that these networks can seed non-linear objects with masses $M\geq 10^{10}  M_\odot$ earlier than expected in standard structure formation models, potentially contributing to a mass excess at redshifts $7\leq z\leq 9$, as recently suggested by JWST data. We have demonstrated, in particular, that a biased domain wall network with a tension of $\sigma \sim (65\ {\rm MeV})^3$ and decaying around $z_{\rm bias}\sim 10^4$ would reproduce the observations described by \citet{Labb__2023}, as it would produce the correct abundance of objects in the mass bins $[10^{9.23}, 10^{10.4}] M_\odot$ and $[10^{10.5},10^{11}] M_\odot$.

Note however that these results should, at this stage, be regarded as preliminary. On the one hand, there are limitations in our modelling of domain wall seeded structure formation. For instance, we have assumed that domain walls are, to a good approximation, spherical (and thus generate spherically symmetric perturbations),  which means that we are calculating the maximal masses these walls could generate. In general, walls are expected to have irregular shapes, which causes them to lose energy both through scalar and gravitational radiation, lowering the energy available upon collapse, and influencing the size of the generated perturbation. On the other hand, there are uncertainties related to the significance of the observations reported in~\citet{Labb__2023}. Domain wall networks are homogeneous and isotropic on sufficiently large scales and we would therefore expect the distribution of domain-wall-seeded structures to be as well. In our analysis, we have assumed that the region observed in~\citet{Labb__2023} is typical and, consequently, that on average we should find the same distribution of objects in every patch of the sky with the same area as the field considered. If we relax this assumption and assume, for instance, that there is only one object with a mass in the range $[10^{10.5},10^{11}] M_\odot$ within an area ten times larger than the survey field, this would reduce the required wall tension by about one order of magnitude. So, definitely establishing the viability of our model and accurately reconstructing the properties of the domain walls would require a detailed survey of the abundance and distribution of large mass objects at these redshifts over larger portions of the sky. Nevertheless, despite these limitations, this analysis does serve as a proof of concept, as it demonstrates that biased domain walls may indeed play a relevant role in structure formation.

It is interesting to note that such a biased domain wall network would evade detection with current and upcoming gravitational wave detectors. The stochastic gravitational wave background generated by domain wall networks is dominated by the last gravitational waves emitted before the full decay of the network. Their spectrum is highly peaked at a frequency $f\sim 10^{-19} (1+z_{\rm bias})/\alpha\,\,{\rm Hz}$~\cite{Gruber:2024jtc,Gruber:2024pqh} (where $\alpha<1$ is a parameter introduced to characterize the typical scale of gravitational wave emission that is generally not expected to be much smaller than unity~\cite{Hiramatsu:2013qaa,Dunsky:2024zdo,Ferreira:2024eru}) and its amplitude quickly decreases as one moves away from this peak. For $z_{\rm bias}\ll 10^{10}$, this peak would be at frequencies that are lower than those Pulsar Timing Arrays are sensitive to and, since these are the lowest frequencies we can currently probe, the stochastic gravitational wave spectrum would be out of our reach. Studying the imprints of biased domain walls on structure formation may then provide us with a novel way to probe domain wall scenarios that are currently not accessible to us, allowing us to uncover new regions of the $(\sigma,z_{\rm bias})$-parameter range.

\begin{acknowledgements} 
The authors thank Ana Paulino-Afonso for enlightening discussions about the JWST results and Dr. Paul Tol for the colorblind-friendly color scheme used in this paper.  C.~W. is supported by FCT - Fundação
para a Ciência e a Tecnologia (https://ror.org/00snfqn58) through the PhD fellowship with reference UI/BD/154758/2023 and DOI https://doi.org/10.54499/UI/BD/154758/2023.~This work was also supported by FCT through the research grants UIDB/04434/2020 and UIDP/04434/2020. For the purpose of Open Access, the authors have applied a CC-BY public copyright license to any Author's Accepted Manuscript (AAM) version arising from this submission.
\end{acknowledgements}

\bibliography{ref}

\end{document}